\documentclass[final,authoryear]{elsarticle}

\usepackage{amssymb}
\usepackage{latexsym}

\usepackage{amsmath}
\usepackage{bbm}  
\usepackage{algorithm,algorithmic}
\usepackage{booktabs}
\usepackage{makecell}
\usepackage{url}
\usepackage{xcolor}

\usepackage{hyperref}

\definecolor{newcolor}{rgb}{.8,.349,.1}

\journal{Medical Image Analysis}

\begin{document}


\begin{frontmatter}

\title{Spatially Regularized Parametric Map Reconstruction for Fast Magnetic Resonance Fingerprinting\tnoteref{tn1}}
\tnotetext[tn1]{F. Balsiger and M. Reyes designed, developed, and evaluated the deep learning component of the proposed solution. B. Marty designed the MRF T1-FF sequence and developed the dictionary matching reconstruction pipeline. B. Marty and P. G. Carlier provided the patient dataset. All authors provided critical feedback and helped to shape the research and manuscript.}

\author[1,2,3,4]{Fabian Balsiger\corref{cor1}}
\cortext[cor1]{Corresponding author.}
\ead{fabian.balsiger@artorg.unibe.ch}

\author[1,2]{Alain Jungo}
\author[5,6]{Olivier Scheidegger}
\author[3,4]{Pierre~G. Carlier}
\author[1,2]{Mauricio Reyes\fnref{fn1}}
\author[3,4]{Benjamin Marty\fnref{fn1}}
\fntext[fn1]{B. Marty and M. Reyes share senior authorship.}

\address[1]{ARTORG Center for Biomedical Engineering Research, University of Bern, Bern, Switzerland}
\address[2]{Insel Data Science Center, Inselspital, Bern University Hospital, University of Bern, Bern, Switzerland}
\address[3]{NMR Laboratory, Institute of Myology, Neuromuscular Investigation Center, Paris, France}
\address[4]{NMR Laboratory, CEA, DRF, IBFJ, MIRCen, Paris, France}
\address[5]{Department of Neurology, Inselspital, Bern University Hospital, University of Bern, Bern, Switzerland}
\address[6]{Support Center for Advanced Neuroimaging (SCAN), Institute for Diagnostic and Interventional Neuroradiology, Inselspital, Bern University Hospital, University of Bern, Bern, Switzerland}

\begin{abstract}
Magnetic resonance fingerprinting (MRF) provides a unique concept for simultaneous and fast acquisition of multiple quantitative MR parameters. Despite acquisition efficiency, adoption of MRF into the clinics is hindered by its dictionary matching-based reconstruction, which is computationally demanding and lacks scalability. Here, we propose a convolutional neural network-based reconstruction, which enables both accurate and fast reconstruction of parametric maps, and is adaptable based on the needs of spatial regularization and the capacity for the reconstruction. We evaluated the method using MRF T1-FF, an MRF sequence for T1 relaxation time of water (T1\textsubscript{H2O}) and fat fraction (FF) mapping. We demonstrate the method's performance on a highly heterogeneous dataset consisting of 164 patients with various neuromuscular diseases imaged at thighs and legs. We empirically show the benefit of incorporating spatial regularization during the reconstruction and demonstrate that the method learns meaningful features from MR physics perspective. Further, we investigate the ability of the method to handle highly heterogeneous morphometric variations and its generalization to anatomical regions unseen during training. The obtained results outperform the state-of-the-art in deep learning-based MRF reconstruction. The method achieved normalized root mean squared errors of 0.048 $\pm$ 0.011 for T1\textsubscript{H2O} maps and 0.027 $\pm$ 0.004 for FF maps when compared to the dictionary matching in a test set of 50 patients. Coupled with fast MRF sequences, the proposed method has the potential of enabling multiparametric MR imaging in clinically feasible time.
\end{abstract}

\begin{keyword}
Magnetic resonance fingerprinting\sep convolutional neural network\sep quantitative magnetic resonance imaging\sep image reconstruction
\end{keyword}

\end{frontmatter}

\section{Introduction}

Magnetic resonance fingerprinting~\citep{Ma2013} (MRF) is a concept for simultaneous and fast acquisition of multiple quantitative MR parameters. The MR acquisition relies on temporal variations of MR sequence parameters usually combined with high \textit{k}-space under-sampling. As a result, a time-series of weighted MR images is acquired, where each tissue has a unique MR signal evolution - or fingerprint. Such fingerprints can be simulated, e.g., by Bloch equations, and a dictionary of expected fingerprints can be built. During image reconstruction, the acquired fingerprints are matched to this dictionary of simulated fingerprints with known MR parameters. The highest correlated fingerprint in the dictionary yields the MR parameters at the given voxel. By repeating this process for all voxels, parametric maps are reconstructed.

MRF has the potential for clinically feasible multiparametric MR imaging, and could enable objective evaluation and comparison for a wide variety of clinical applications~\citep{Poorman2019}. However, whilst the MRF acquisition is fast, the dictionary matching reconstruction is computationally demanding and lacks scalability as the problem worsens exponentially with the number of reconstructed MR parameters. For instance, as reported in~\citet{Marty2019a}, the reconstruction of five parametric maps can require minutes to hours depending on the implementation and computational hardware. This long reconstruction is mainly attributed to the large dictionary with approximately 9 million simulated fingerprints for five parametric maps. The reconstruction time will especially proof problematic when acquiring large data sets with many slices in clinically settings. Additionally, the dictionary matching results in discretized parametric maps, which might be undesirable considering continuous-valued parametric maps using gold-standard MR sequences. Therefore, the dictionary matching represents currently a drawback of MRF, which makes a routine clinical application of MRF potentially inappropriate, and calls for accurate and fast reconstruction alternatives.

Several methods attempting to improve the MRF reconstruction have been proposed lately. Acceleration of the dictionary matching~\citep{McGivney2014,Cauley2015,Gomez2016}, iterative reconstruction~\citep{Davies2014,Pierre2016}, and low-rank approximations~\citep{Mazor2018,Asslander2018,Zhao2018,LimadaCruz2019} were proposed for MRF reconstruction. While some of these methods are promising, both further acceleration of the reconstruction process and continuous values in the parametric maps, as opposed to the discretely sampled dictionary-based reconstruction, are highly desired. Promising in these regards are deep learning-based methods, which offer near real-time reconstructions and produce parametric maps with continuous values.

Deep learning-based methods for MRF reconstruction are versatile but can coarsely be classified into fingerprint-wise reconstruction and spatially regularizing reconstruction. Fingerprint-wise methods feed single fingerprints into a neural network that regresses the MR parameters of interest. Such methods can directly be trained with the entries of the dictionaries but also with the fingerprints of acquired MRF data. \citet{Hoppe2017} proposed a neural network with three 1-D convolutions followed by a fully-connected layer for fingerprint-wise regression of MR parameters. Similarly, \citet{Cohen2018} relied on a solely fully-connected architecture with two hidden layers. A very similar fully-connected architecture was proposed by \citet{Golbabaee2019} with three hidden layers. A more complex architecture was proposed by \citet{Song2019} using residual learning combined with attention mechanisms. Also, in the context of fingerprint-wise reconstruction, \citet{Virtue2017} investigated the complex-valued nature of MRF data by using a complex-valued fully-connected neural network. Recently, \citet{Oksuz2019} used recurrent neural networks (RNNs), where the inputs to the RNN were also fingerprints, followed by a fully-connected layer that regressed the MR parameters. The hypothesis that information between neighboring fingerprints, especially in highly undersampled MRF, could benefit the reconstruction lead researchers exploring spatially regularizing methods. Here, a neighborhood of fingerprints is fed to a neural network that regresses a spatial patch in the parametric maps. These methods are usually trained on acquired MRF data because spatial data, i.e., image slices, is required. \citet{Balsiger2018b} proposed a convolutional neural network (CNN) regressing MR parameters from a neighborhood of 5 $\times$ 5 fingerprints. The work of \citet{Fang2018} used a U-Net architecture, considering a neighborhood of fingerprints, for the estimation of parametric maps. Their follow-up work \citep{Fang2019} additionally proposed to use a feature extraction module that reduces the dimensionality of fingerprints prior to feeding patches of fingerprints to the U-Net. In total, 54 $\times$ 54 fingerprints are used for spatial regularization. Clearly, spatial regularization works superior to fingerprint-wise reconstruction, however, we argue that such strong spatial regularization, and especially spatial pooling and upsampling operations, as in \citet{Fang2018,Fang2019} is not needed.

We hypothesize that the reconstruction performance is mainly dependent on 1) the extent of spatial regularization and 2) the capacity of the CNN. Motivated by the vast amount of MRF sequences and their differences in fingerprint dimensionality \citep{Poorman2019}, we believe that a CNN architecture, which is adaptable to the specific needs of different MRF sequences, is necessary. To prove the hypothesis, we propose an algorithm that builds the CNN architecture based on the needs of spatial regularization and capacity. The backbone of the CNN was presented in our conference contribution \citep{Balsiger2019a}, which we extend by the algorithm making the CNN adaptable and possibly useful for different types of MRF sequences. We evaluated the CNN's performance on a large (n=164) and highly heterogenous patient dataset, and compared the method to four existing deep learning-based methods proposed by \citet{Cohen2018,Hoppe2017,Oksuz2019,Fang2019}. As in \citet{Balsiger2019a}, we empirically show the benefit of incorporating spatial regularization during the reconstruction and demonstrate that the CNN learns meaningful features from MR physics perspective. Additionally, we investigated the ability of the CNN to handle highly heterogeneous morphometric variations and its generalization to anatomical regions unseen during training.

\section{Materials and Methods}

\subsection{MRF Acquisition and Dictionary Matching Reconstruction}
\label{sec:mrfacquisition}

We used MRF T1-FF~\citep{Marty2019a}, an MRF sequence for T1 relaxation time (T1) and fat fraction (FF) mapping in fatty infiltrated tissues. Fatty infiltration occurs for instance in neuromuscular diseases, where muscle cells are irreversible replaced by fat resulting in loss of muscle strength. In such cases, the FF in muscles is often quantified as biomarker for disease severity \citep{Carlier2016,Paoletti2019}. Additionally, increased T1 can be found in diseased muscle tissue \citep{Marty2019}, which might reflect disease activity. However, in the presence of fatty infiltration, the T1 quantification can be biased by the fat, necessitating the separation of the water and fat pools resulting in T1 of water (T1\textsubscript{H2O}) and T1 of fat (T1\textsubscript{fat}). The MRF T1-FF sequence is specifically developed for such a separation of water and fat, and, is therefore capable of quantifying FF, T1\textsubscript{H2O}, and T1\textsubscript{fat}. The acquisition of MRF T1-FF consisted of a 1400 radial spokes FLASH echo train following the golden angle scheme after non-selective inversion. Echo time, repetition time, and nominal flip angle were varied during the echo train. The field of view was set to 350~mm~$\times$~350~mm with a voxel size of 1.0~mm~$\times$~1.0~mm~$\times$~8.0~mm, and five slices were acquired within a total acquisition time of 50~seconds. All acquisitions were performed on a 3 tesla Siemens MAGNETOM Prisma\textsuperscript{fit} scanner (Siemens Healthineers, Erlangen, Germany) using a set of 18-channel flexible phase array coils, combined with a 48-channel spine coil.

Five parametric maps were reconstructed after the MRF T1-FF acquisition: FF, T1\textsubscript{H2O}, T1\textsubscript{fat}, and additionally the two confounding factors static magnetic field inhomogeneity ($\Delta$f) and flip angle efficacy (B1). First, the acquired data was transformed to image space using the non-uniform fast Fourier transform (NUFFT)~\citep{Fessler2003a} with eight spokes per temporal frame, resulting in a highly undersampled time series of 175 temporal frames (acceleration factor of 68.7). Second, dictionary matching was conducted using a dictionary simulated by Bloch equations with $(0{:}0.05{:}1)$ for FF, $(550{:}10{:}1600,1650{:}50{:}2000)$~ms for T1\textsubscript{H2O}, $(225{:}25{:}400)$~ms for T1\textsubscript{fat}, $(-120{:}10{:}120)$~Hz for $\Delta$f, and $(0.3{:}0.05{:}1)$ for B1 (start:increment:stop). Despite fast group matching~\citep{Cauley2015} and dictionary compression~\citep{McGivney2014}, the dictionary matching still required approximately 5~hours using standard desktop computer hardware (2.6~GHz Intel Xenon E5-2630, 48~GB memory) due to the large number of MR parameter combinations. In summary, the temporal dimensionality of the fingerprints of MRF T1-FF is 175, the spatial dimensionality is 350 $\times$ 350, and five parametric maps are quantified.

\subsection{Conceptual Formulation}

\begin{figure*}[!t]
\centering
\includegraphics[width=1.0\textwidth]{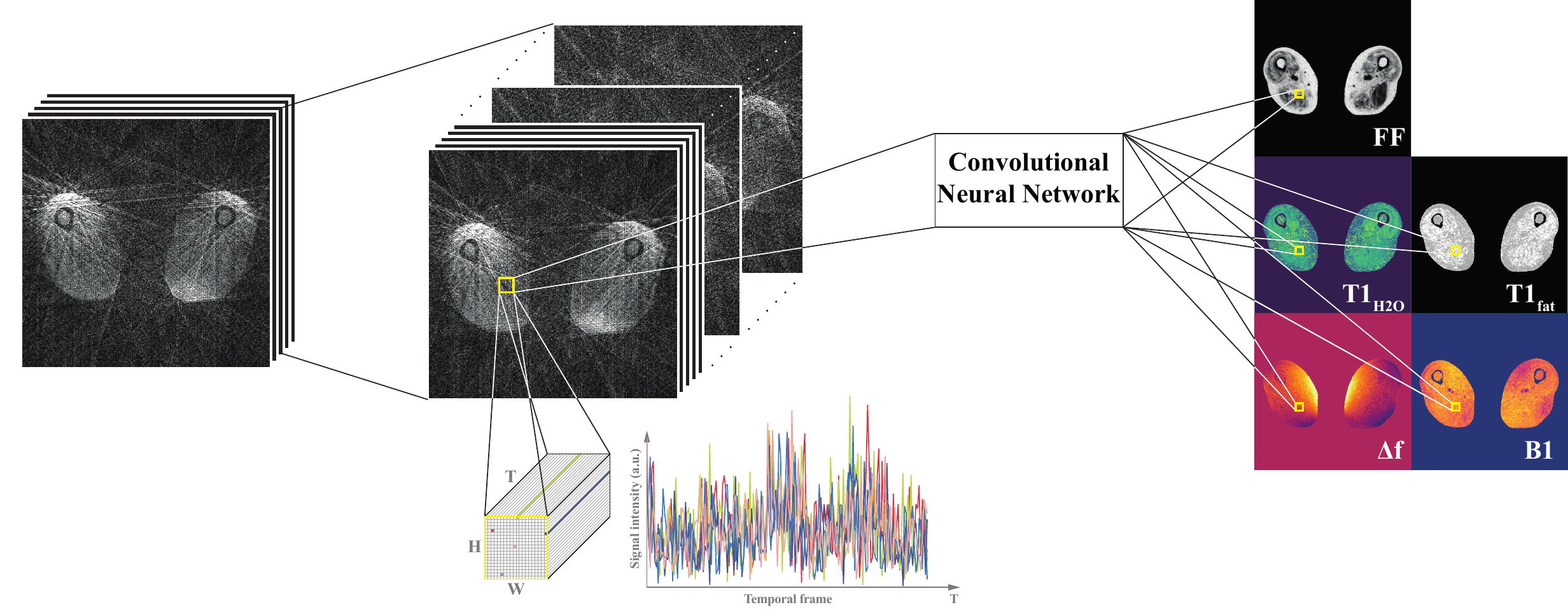}
\caption{Schematic overview of the proposed MRF reconstruction. Patches of H~$\times$~W fingerprints with T temporal frames are extracted from MRF image slices and fed to a CNN, which simultaneously predicts all parametric maps. We used MRF T1-FF~\citep{Marty2019a}, an MRF sequence to image diseased skeletal muscle. The parametric maps show the thighs of a 69 years old male patient with inclusion body myositis. FF: fat fraction, T1\textsubscript{H2O}: T1 relaxation time of water, T1\textsubscript{fat}: T1 relaxation time of fat, $\Delta$f: static magnetic field inhomogeneity, B1: flip angle efficacy.}
\label{fig:overview}
\end{figure*}

The hypothesis leading to the design of the proposed CNN architecture is that the reconstruction performance depends on 1) the CNN's receptive field and 2) the CNN's capacity. On one hand, the receptive field determines the number of neighboring fingerprints the CNN will use to predict the value of the parametric maps of the central fingerprint. We argue that it is, especially in the case of highly undersampled MRF, beneficial to leverage the spatial correlation among fingerprints but this spatial correlation is limited to a certain extent due to different tissue properties yielding different fingerprints, especially in lesions. Technically speaking, the extent of spatial correlation limits the number of spatial convolutions, i.e., convolutions with kernel sizes larger or equal than $3 \times 3$. On the other hand, the reconstruction performance will also be determined by the capacity of the CNN, i.e., the number of learnable parameters or number of convolutional filter weights. A certain capacity is required to extract features that cover the space of possible input fingerprints. Similar as for the receptive field, an appropriate capacity is especially needed when dealing with multiple parametric maps and diverse tissue properties. Having both factors adaptable by an algorithm allows specific tailoring of the CNN-based reconstruction to the MRF sequence at hand.

The schematic overview of the proposed MRF reconstruction is shown in Fig.~\ref{fig:overview}. Let us consider a 2-D+time MRF image slice $I \in \mathbb{C}^{H \times W \times T}$ after NUFFT in the image space with matrix size $H \times W$ and $T$ temporal frames. We aim to find the mapping $\mathcal{M}: I \rightarrow Q$ from the MRF image space to $M$ parametric maps $Q \in \mathbb{R}^{H \times W \times M}$. To learn this mapping, we use a 2-D CNN parametrized by its convolutional filter weights $\theta$. The CNN processes the MRF data patch-wise and treats the temporal frames as channels. Therefore, the CNN is trained to learn the mapping $f: I_P \rightarrow Q_P$, and estimates the parametric maps by

\begin{equation}
\label{eq:mapping}
\hat{Q}_P = f(I_P; \theta),
\end{equation}

\noindent where $f$ the non-linear mapping of the CNN parametrized by $\theta$, and $I_P \in \mathbb{C}^{IP_H \times IP_W \times T} \subset I$ and $Q_P \in \mathbb{R}^{QP_H \times QP_W \times M} \subset Q$ are patches extracted from the 2-D+time MRF image slice and the parametric maps. The CNN reconstructs non-overlapping patches of the parametric maps with size of $QP_H \times QP_W$\footnote{The patch-wise processing is mainly motivated by graphics processing unit (GPU) memory limitations, in this study 12~GB.}. Due to the use of valid convolutions, the input patch size is larger and determined with the CNN's receptive field $R$ by $IP_H \times IP_W = QP_H + R - 1 \times QP_W + R - 1$. Meaning, the input patches are larger than the reconstructed patches at the output of the CNN. We chose the output patch size to be of dimension $QP_H \times QP_W = 32 \times 32$ and the input patch dimension was $IP_H \times IP_W = 46 \times 46$. Therefore, we used a receptive field of $15 \times 15$, i.e., $R = 15$. Further for MRF T1-FF, $H = W = 350$, $T = 175$, and $M = 5$ (FF, T1\textsubscript{H2O}, T1\textsubscript{fat}, $\Delta$f, B1).

\subsection{CNN Architecture}
\begin{figure*}[!t]
\centering
\includegraphics[width=1.0\textwidth]{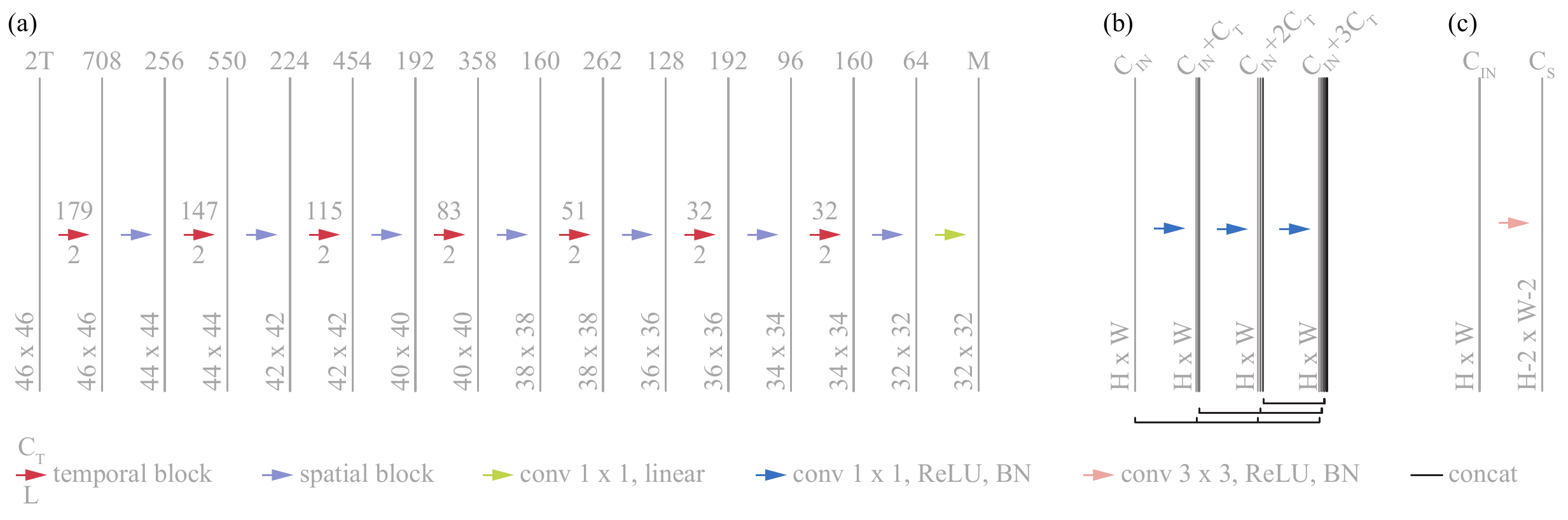}
\caption{The proposed CNN for MRF reconstruction with a receptive field of $15 \times 15$ ($R = 15$). (a) the architecture for MRF T1-FF and its (b) temporal (here $L = 3$) and (c) spatial blocks. The bars indicate feature maps with the number of channels indicated on the top and the spatial size indicated on the lower left. T: number of temporal frames, M: number of parametric maps, BN: batch normalization, ReLU: rectified linear unit, L: number of layers in a temporal block, $C_T$/$C_S$: number of channels in a temporal/spatial block, $C_{IN}$: number of input channels, H~$\times$~W: feature map size.}
\label{fig:architecture}
\end{figure*}

The CNN architecture consists of temporal and spatial blocks, which are interleaved within the architecture as shown in Fig.~\ref{fig:architecture}a. The temporal blocks extract temporal features from fingerprints while maintaining the receptive field of the CNN. The spatial blocks extract spatially correlated features and increase the receptive field of the CNN. By appropriately setting the number of channels in the temporal and spatial blocks, the capacity can be adjusted. The interleaved blocks are followed by a 1~$\times$~1 convolution with linear activation function and $M$ channels for predicting $Q_P$. Input to the CNN are real-valued $I_P$ with real and imaginary parts concatenated as $2T$ channels. A temporal block (Fig.~\ref{fig:architecture}b) consists of $1 \times 1$ convolutions to not increase the receptive field and follows the design of dense blocks~\citep{Huang2017}. They are composed of $L$ layers, and each of them is a sequence of 1~$\times$~1 convolution, rectified linear unit (ReLU) activation function~\citep{Glorot2011}, and batch normalization (BN)~\citep{Ioffe2015}. All convolutions have the same number of $C_T$ filters (growth rate), and the feature maps of the preceding layers are concatenated before the next layer to reuse features and facilitate the gradient flow~\citep{Huang2017}. A spatial block (Fig.~\ref{fig:architecture}c) extracts features from neighboring fingerprints, and, therefore, increases the CNN's receptive field. It consists of a valid 3~$\times$~3 convolution with $C_S$ filters followed by ReLU activation function, and BN. All convolutions in the CNN are performed with a stride of 1. In principle, the temporal blocks extract a high number of channels, from which the spatial blocks then extract spatially correlated features with an even higher number of parameters (factor 9 due to 3~$\times$~3 convolution). This increases both the receptive field and the capacity of the CNN.

\subsection{Algorithm}
We propose Algorithm~\ref{algo:architecture} to parametrize the temporal and spatial blocks such that the receptive field and the capacity of the CNN are as desired for the MRF sequence to reconstruct. Inputs to the algorithm are the receptive field $R$, the number of parameters $N_{P}$, i.e., the number of learnable weights of all convolutional kernels in the CNN, and the number of non-linearities $N_{L}$, i.e., the number of ReLU activation functions\footnote{Note that the number of non-linearities are equal to the number of convolutions in the temporal and spatial blocks because each convolution is directly followed by a ReLU activation function.}. The number $N_B = (R - 1) / 2$ of temporal and spatial blocks are determined by the receptive field. The number of channels of the spatial blocks $C_S$ are chosen such that it gradually decreases down to $C_{S_{stop}}$, in steps of $C_{S_{dec}}$, before the last convolutional filter with linear activation, i.e.,

\begin{equation}
\label{eq:spatialblock}
C_S = \left ( i C_{S_{dec}} + C_{S_{stop}} \right )_{i=N_B-1}^{0},
\end{equation}

\noindent where $(\cdot)$ denotes a sequence (e.g., $C_S = (160, 128, 96, 64)$ for $N_B = 4$, $C_{S_{stop}} = 64$, and $C_{S_{dec}} = 32$). The number of layers $L$ in each of the temporal blocks are determined by 

\begin{equation}
\label{eq:nonlinearities}
L = \left ( \left \lfloor \dfrac{N_{L_T}}{N_B} \right \rfloor + \mathbbm{1}_{(N_{L_T} \mod N_B) \leq i} \right )_{i=1}^{N_B},
\end{equation}

\noindent where $N_{L_T} = N_L - N_B$ is the remaining number of non-linearities, which corresponds to the total number of convolutional layers in all temporal blocks. The indicator function $\mathbbm{1}$ returns $1$ if the statements is true and $0$ otherwise, and it allows uneven distribution of the convolutional layers among the temporal blocks (e.g., $L = (3, 3, 2, 2)$ for $N_B = 4$ and $N_{L_T} = 10$). Given that the number of filters of the temporal blocks $C_T$ gradually decrease down to $C_{T_{stop}}$, in steps of $C_{T_{dec}}$, an ideal number of channels in the first temporal block $C_{T_{start}}$ can be calculated such that the total number of learnable convolutional parameters are as close as possible to the desired number of learnable parameters $N_{P}$\footnote{Calculating the number of parameters $N$ of a convolution is straightforward, i.e., $K^2 * C_{in} * C_{out} + C_{out}$ for a convolution with kernel size $K \times K$, $C_{in}$ input channels, and $C_{out}$ output channels.}. Therefore, the desired capacity of the CNN can be matched as close as possible. $C_T$ is calculated by

\begin{equation}
\label{eq:temporalblock}
C_T = \left ( g(i) \right )_{i=0}^{N_B-1}, 
\end{equation}

\noindent with 

\begin{equation}
\label{eq:temporalblock2}
g(x)= 
\begin{cases}
    C_{T_{start}} - x C_{T_{dec}} , & \text{if } C_{T_{start}} - x C_{T_{dec}} \geq C_{T_{stop}}\\
    C_{T_{stop}},              & \text{otherwise}
\end{cases},
\end{equation}

\noindent (e.g., $C_T = (80, 48, 32, 32)$ for $N_B = 4$, $C_{T_{start}} = 80$, $C_{T_{stop}} = 32$, and $C_{T_{dec}} = 32$). As we will analyze in Section~\ref{sec:spatial}, the optimal receptive field of the CNN for the MRF T1-FF sequence is 15~$\times$~15 with approximately 5~million parameters. Therefore, the architecture consists of seven temporal and spatial blocks. We empirically chose the hyperparameters $N_{L} = 21$, $C_{S_{stop}} = 64$, $C_{S_{dec}} = 32$, $C_{T_{stop}} = 32$, and $C_{T_{dec}} = 32$, resulting in number of layers $L = (2, 2, 2, 2, 2, 2, 2, 2)$, channels of the temporal blocks $C_T = (179, 147, 115, 83, 51, 32, 32)$, i.e. $C_{T_{start}} = 179$, and channels of the spatial blocks $C_S = (256, 224, 192, 160, 128, 96, 64)$.

\begin{algorithm}[H]
 \caption{Algorithm for CNN architecture building.}
 \label{algo:architecture}
 \begin{algorithmic}[1]
 \renewcommand{\algorithmicrequire}{\textbf{Input:}}
 \renewcommand{\algorithmicensure}{\textbf{Output:}}
 \REQUIRE $R$, $N_P$, $N_L$, $C_{S_{stop}}$, $C_{S_{dec}}$, $C_{T_{stop}}$, $C_{T_{dec}}$
 \ENSURE $L, C_T, C_S$
  \STATE $N_B \gets (R - 1) / 2$, $N_{L_T} \gets N_L - N_B$
  \STATE Calculate $C_S$ with Eq.~\ref{eq:spatialblock}
  \STATE Calculate $L$ with Eq.~\ref{eq:nonlinearities}
 \STATE $C_{T_{start}} \gets 1$, $N_{P_{best}} \gets \infty$
  \LOOP
  \STATE Calculate $C_T$ with Eq.~\ref{eq:temporalblock}
  \STATE $N_P{_{current}} \gets calculate\_parameters(L, C_{T}, C_S)$
  \IF {$ \lvert N_P - N_P{_{current}} \rvert < N_{P_{best}}$}
  \STATE $N_{P_{best}} \gets N_P{_{current}}$
  \STATE $C_{T_{start}} \gets C_{T_{start}} + 1$
  \ELSIF{$\lvert N_P - N_P{_{current}} \rvert > N_{P_{best}}$}
  \STATE $C_{T_{start}} \gets C_{T_{start}} - 1$
  \STATE Calculate $C_T$ with Eq.~\ref{eq:temporalblock}
  \RETURN $L, C_T, C_S$
  \ELSE
  \RETURN $L, C_T, C_S$
  \ENDIF
  \ENDLOOP
 \end{algorithmic} 
\end{algorithm}

\subsection{CNN Training and Implementation}

The CNN was trained for 75 epochs with a batch size of 20 randomly selected patches, which we empirically found to be sufficient. The Adam optimizer~\citep{Kingma2015} was used to minimize a mean squared error (MSE) loss ($\ell$) with a learning rate of $0.001$, $\beta_1 = 0.9$, and $\beta_2 = 0.999$. Therefore, the learning objective was

\begin{equation}
\label{eq:optimization}
\min_{\theta} \sum_{i} \ell(f(I_{Pi}; \theta), Q_{Pi}).
\end{equation}

Before training, we normalized the data subject-wise: The MRF image was normalized to zero mean and unit standard deviation along the real and imaginary parts and each parametric map was rescaled to the range $\left[0, 1\right]$ using the minimum and maximum values in the dictionary (see Section~\ref{sec:mrfacquisition}). After inference, the predicted parametric maps were rescaled back to the original dictionary range. The CNN was implemented in TensorFlow 1.10.0 (Google Inc., Mountain View, CA, U.S.) with Python 3.6.7 (Python Software Foundation, Wilmington, DA, U.S.). The training was performed with an NVIDIA TITAN Xp (Nvidia Corporation, Santa Clara, CA, U.S.). For reproducibility, the source code is available online\footnote{https://github.com/fabianbalsiger/mrf-reconstruction-media2020}. Further investigation of some architecture hyperparameters can be found in Section~1 of the supplementary material.

\subsection{Evaluation and Comparison}
\label{sec:evaluation}

\begin{table}[!t]
\renewcommand{\arraystretch}{1.1}
\caption{Clinical and demographic information of the dataset and its distribution into training, validation, and testing splits. Values are given as mean age $\pm$ standard deviation and number of total subjects / number of male subjects / number of thigh images. BMD: Becker muscular distropy, DMD: Duchenne muscular distropy, IBM: Inclusion body myositis.}
\label{tab:materials}
\centering
\begin{tabular}{lcccc}
\toprule
 & & \multicolumn{3}{c}{Split} \\
\cmidrule{3-5}
Disease & Overall  & Training &  Validation &  Testing \\
\midrule
BMD & 45.4~$\pm$~15.0 & 45.7~$\pm$~16.9 & 64.0~$\pm$~0.0 & 42.3~$\pm$~11.3 \\
       & 19 / 19 / 19 & 11 / 11 / 11 & 1 / 1 / 1 & 7 / 7 / 7 \\
DMD & 12.4~$\pm$~2.2 & 12.5~$\pm$~2.3 & 12.8~$\pm$~2.5 & 11.9~$\pm$~2.0 \\
      & 25 / 25 / 12 & 14 / 14 / 8 & 4 / 4 / 1 & 7 / 7 / 3 \\
IBM & 64.5~$\pm$~9.6 & 65.3~$\pm$~10.4 & 66.3~$\pm$~10.3 & 62.6~$\pm$~8.3 \\ 
& 70 / 30 / 34 & 37 / 16 / 13 & 9 / 4 / 7 & 24 / 10 / 14 \\
Other & 46.9~$\pm$~14.4 & 49.2~$\pm$~14.6 & 45.0~$\pm$~16.0 & 41.7~$\pm$~12.8 \\
 & 50 / 19 / 26 & 32 / 8 / 15 & 6 / 6 / 4 & 12 / 5 / 7 \\
\midrule
Overall & 49.0~$\pm$~21.0 & 49.7~$\pm$~21.2 & 49.1~$\pm$~23.4 & 47.6~$\pm$~19.8 \\
        & 164 / 93 / 91 & 94 / 49 / 47 & 20 / 15 / 13 & 50 / 29 / 31 \\
\bottomrule
\end{tabular}
\end{table}

We evaluated the performance of the proposed method on a clinical dataset consisting of 164 patients with various neuromuscular diseases (NMDs). The dataset is highly heterogeneous due to the variable phenotypic appearance of lesions in NMDs, and further comprises thigh and leg images. To evaluate the robustness of the methods, we purposely did not apply any stratification regarding disease type, patient sex, patient age, or anatomical region when splitting the dataset into training, validation, and testing splits (n=94/20/50). Table~\ref{tab:materials} summarizes clinical and demographic characteristics of the dataset. Multimedia files characterizing the heterogeneity of the dataset can be found online as supplementary material.

The dictionary matching reconstruction served as reference for the parametric maps. Quantitative analysis between the dictionary matching and the predicted parametric maps was done according to~\citet{Zbontar2018}. The normalized root mean squared error (NRMSE), the peak signal-to-noise ratio (PSNR), and the structural similarity index measure (SSIM)~\citep{Wang2004} were calculated at the image level. Due to the quantitative nature of parametric maps, we provide further quantitative analysis based on the coefficient of determination (R\textsuperscript{2}), scatter plots, and Bland-Altman analysis. To this end, we manually segmented regions of interest (ROIs) lying within the major muscles of each subject. The ROIs allowed calculating the mean parametric value within each ROI of each image slice. Then a linear regression between the mean ROI values of the dictionary matching and predicted parametric maps quantified the agreement between the methods. For all evaluation, background voxels (air) were excluded based on an automatically segmented mask generated by thresholding an anatomical image obtained from the MRF image space series (pseudo out-of-phase image~\citep{Marty2019a}). Further, voxels and ROIs with a FF higher than 0.7 were excluded from the evaluation of NRMSE, PSNR, and R\textsuperscript{2} of the T1\textsubscript{H2O} map reconstruction due to low confidence of T1\textsubscript{H2O} at high FF~\citep{Marty2019a}. For the SSIM, we used a window size of 7~$\times$~7, $K_1 = 0.01$, $K_2 = 0.03$, and $L$ was set to the maximum value of the parametric map.

We compared the proposed method to four other deep learning-based MRF reconstruction, which can be grouped into methods working fingerprint-wise and a method considering spatial neighborhoods of fingerprints. The fingerprint-wise methods comprise \citet{Cohen2018}, a neural network with two hidden fully-connected layers, \citet{Hoppe2017}, a CNN with four 1-D convolutional layers, and \citet{Oksuz2019}, a recurrent neural network (RNN) using a gated recurrent unit with 100 neurons. The spatial method proposed by \citet{Fang2019} works patch-wise using an U-Net-like CNN with pooling operations resulting in a receptive field of 54~$\times$~54. We implemented all competing methods as described in the papers due to lack of publicly available code. The input and output dimensions were adapted for MRF T1-FF using the complex-valued MRF data as input for all methods. For training, we used the Adam optimizer with a MSE loss as for the proposed method. The batch sizes were set to 100 for the fingerprint-wise methods and the feature extraction module of \citet{Fang2019}, and to 20 for the spatially-constrained quantification module of \citet{Fang2019}. Training was performed for 25 epochs \citep{Cohen2018,Hoppe2017,Oksuz2019} and 75 epochs for \citet{Fang2019}. For each method, the learning rates were chosen from the set $\{0.01, 0.001, 0.0001 \}$ based on the performance on the validation set.

\section{Experiments and Results}

\subsection{Parametric Map Reconstructions}
\begin{figure}[!t]
\centering
\includegraphics[width=0.59\textwidth]{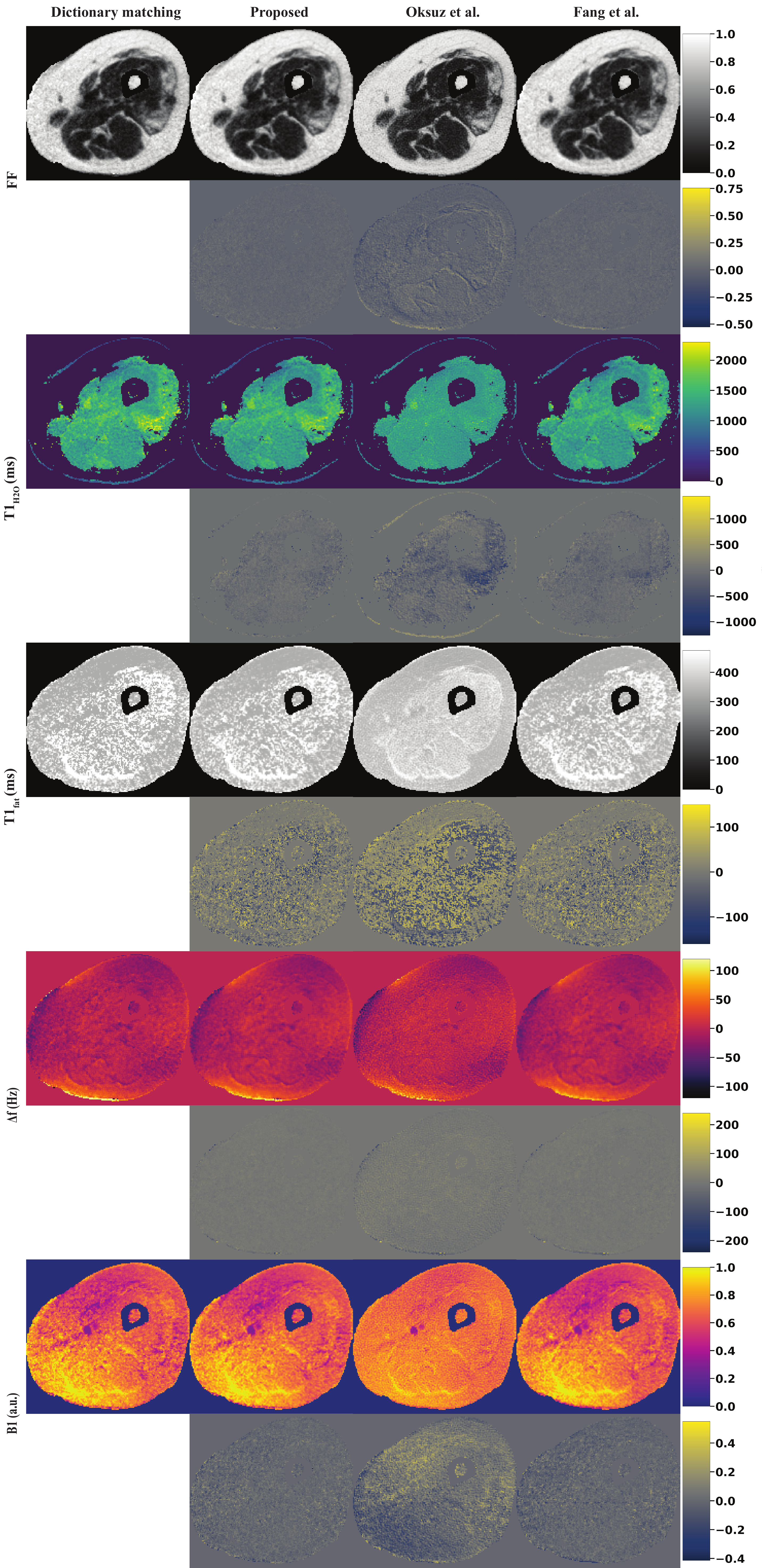}
\caption{Parametric map reconstruction results of a thigh of a 71 years old male patient with inclusion body myositis. Reconstructions of the dictionary matching, the proposed method, \citet{Oksuz2019}, \citet{Fang2019}, and the error (dictionary minus reconstruction) are shown. a.u.: arbitrary unit.}
\label{fig:qualitative}
\end{figure}

Reconstruction results of the dictionary matching, the proposed method, the best fingerprint-wise method~\citep{Oksuz2019}, and the spatial method of \citet{Fang2019} are shown in Fig.~\ref{fig:qualitative}. Visually, the proposed method achieved the best reconstruction results for all parametric maps. Compared to the dictionary matching, all reconstructions appear to be slightly smoothed. \citet{Oksuz2019} resulted in noisier reconstructions and could not capture elevated T1\textsubscript{H2O}. \citet{Fang2019} achieved similar results as the proposed method, but the reconstructions contain artifacts, which are not present for the proposed method. The artifacts possibly originate from the patch-wise processing in combination with padding convolutions, i.e. equal spatial dimension of the input and output of their CNN, resulting in boundary effects. Further, the reconstructions of \citet{Fang2019} appear to be slightly more smooth than the reconstructions of the proposed method. A zoomed-in region with fatty infiltrated muscle and elevated T1\textsubscript{H2O} can be found in Section~2 of the supplementary material, showing that \citet{Oksuz2019} fails to reconstruct elevated T1\textsubscript{H2O} and that the reconstructions of \citet{Fang2019} contain subtle reconstruction artifacts.

\begin{figure}[!t]
\centering
\includegraphics[width=0.62\textwidth]{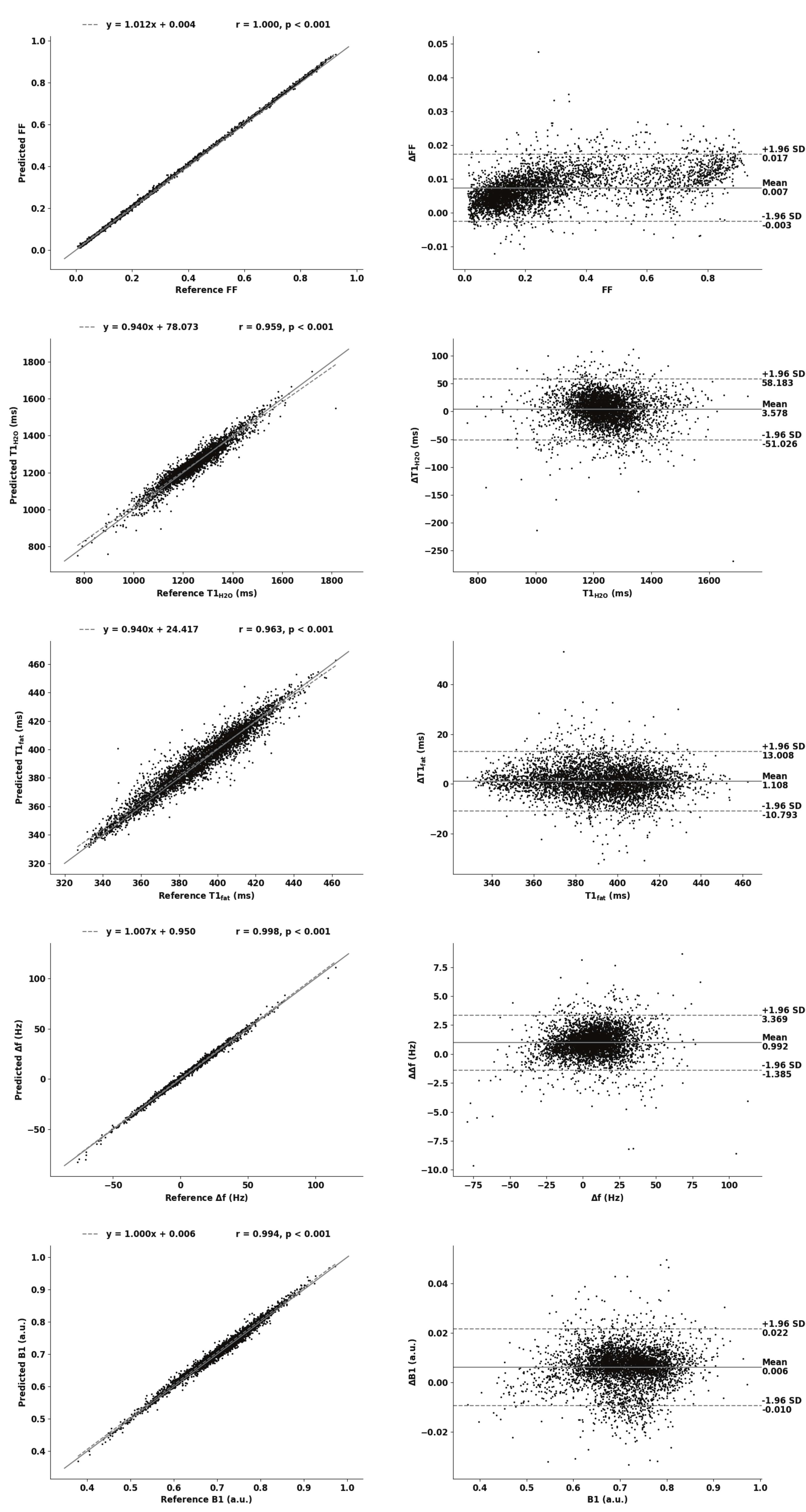}
\caption{Quantitative agreement between the proposed method and the dictionary matching. (left) Scatter and (right) Bland-Altman plots for the five parametric maps where each dot represents the mean value of the parametric map for a manually segmented ROI lying within a major muscle (n=4392 for FF, T1\textsubscript{fat}, $\Delta$f, and B1, and n=3943 for T1\textsubscript{H2O}). For the scatter plots, the solid line indicates x=y and the dashed line indicates the fit of the linear regression. For the Bland-Altman plots, the solid line indicates the mean difference and dashed lines indicate the 95~\% limits of agreement between the dictionary matching and the proposed method.}
\label{fig:quantitative}
\end{figure}

Quantitatively, the proposed method achieved the best reconstruction results for the four metrics (Table~\ref{tab:quantiative}). The parametric maps T1\textsubscript{H2O} and T1\textsubscript{fat} are the most difficult to reconstruct while for FF, $\Delta$f, and B1 the quantitative results are better. The quality of the agreement is further shown by the quantitative analysis of the ROIs in Fig.~\ref{fig:quantitative}. The correlations between the CNN and the dictionary matching reconstruction are very high with the Pearson correlation coefficient $r > 0.95$ (Fig.~\ref{fig:quantitative}, left column). Only for T1\textsubscript{H2O} and T1\textsubscript{fat}, the agreements are slightly decreased with R\textsuperscript{2}s of 0.919 and 0.927. The Bland-Altman plots (Fig.~\ref{fig:quantitative}, right column) show small to no bias for all five parametric maps, and the 95~\% limits of agreement are smaller than the dictionary sampling increment for FF, T1\textsubscript{fat}, $\Delta$f, and B1 (cf. Section~\ref{sec:mrfacquisition}). For T1\textsubscript{H2O}, the agreement between the methods is approximately $\pm$6 sampling steps or $\pm 60~\textnormal{ms}$. Similar plots for the \citet{Oksuz2019} and \citet{Fang2019} can be found in Section~2 of the supplementary material. Reconstructing the parametric maps of one subject (five image slices) required approximately 1~second with the proposed method, which is considerably faster than the dictionary matching requiring up to minutes or even hours depending on the implementation~\citep{McGivney2019}. Our dictionary matching implementation requires approximately 5~hours per subject \citep{Marty2019a}. The compared deep learning-based methods are also in the range of 1~second with the fingerprint-wise methods \citep{Hoppe2017,Cohen2018} being slightly slower than the proposed CNN, followed by the RNN of \citet{Oksuz2019}. The two-stage process of \citet{Fang2019} resulted in the longest reconstruction times.

\begin{table*}[!t]
\tiny
\caption{Quantitative results of the proposed and the compared methods. The metrics normalized root mean squared error (NMRSE), peak signal-to-noise ratio (PSNR), structural similarity index measure (SSIM), and coefficient of determination (R\textsuperscript{2}) were calculated for the five parametric maps.}
\label{tab:quantiative}
\centering
\begin{tabular}{llccccc}
\toprule
& & \multicolumn{5}{c}{Method} \\
\cmidrule{3-7}
Metric & \makecell[l]{Parametric\\ map}  & Proposed & Hoppe et al. & Cohen et al. & Oksuz et al. & Fang et al. \\
\midrule
NRMSE & FF                    & \textbf{0.027~$\pm$~0.004}  &  0.069~$\pm$~0.007  &  0.069~$\pm$~0.006  &  0.063~$\pm$~0.007  &  0.030~$\pm$~0.004 \\
      & T1\textsubscript{H2O} & \textbf{0.048~$\pm$~0.011}  &  0.097~$\pm$~0.021  &  0.100~$\pm$~0.023  &  0.090~$\pm$~0.018  &  0.054~$\pm$~0.012 \\
      & T1\textsubscript{fat} & \textbf{0.212~$\pm$~0.076}  &  0.290~$\pm$~0.095  &  0.294~$\pm$~0.096  &  0.287~$\pm$~0.094  &  0.217~$\pm$~0.077 \\  
      & $\Delta$f                    & \textbf{0.027~$\pm$~0.008}  &  0.062~$\pm$~0.013  &  0.062~$\pm$~0.009  &  0.056~$\pm$~0.008  &  0.030~$\pm$~0.007 \\
      & B1                    & \textbf{0.056~$\pm$~0.009}  &  0.107~$\pm$~0.019  &  0.117~$\pm$~0.020  &  0.107~$\pm$~0.018  &  0.062~$\pm$~0.010 \\
\midrule
PSNR (dB)  & FF               & \textbf{31.6~$\pm$~1.04}  &  23.3~$\pm$~0.95  &  23.4~$\pm$~0.78  &  24.1~$\pm$~0.93  &  30.7~$\pm$~1.02 \\
      & T1\textsubscript{H2O} & \textbf{28.6~$\pm$~1.75}  &  22.4~$\pm$~1.71  &  22.2~$\pm$~1.76  &  23.1~$\pm$~1.54  &  27.6~$\pm$~1.69 \\
      & T1\textsubscript{fat} & \textbf{22.7~$\pm$~1.85}  &  19.9~$\pm$~1.35  &  19.8~$\pm$~1.30  &  20.0~$\pm$~1.38  &  22.5~$\pm$~1.78 \\
      & $\Delta$f                    & \textbf{25.1~$\pm$~2.11}  &  17.7~$\pm$~1.96  &  17.6~$\pm$~1.62  &  18.5~$\pm$~1.63  &  24.1~$\pm$~2.02 \\
      & B1                    & \textbf{27.8~$\pm$~1.06}  &  22.1~$\pm$~1.09  &  21.3~$\pm$~0.96  &  22.2~$\pm$~1.12  &  26.9~$\pm$~1.05 \\
\midrule
SSIM  & FF                     & \textbf{0.984~$\pm$~0.011}   &  0.933~$\pm$~0.039  &  0.934~$\pm$~0.038  &  0.939~$\pm$~0.036  &  0.980~$\pm$~0.014 \\
      & T1\textsubscript{H2O}  &  \textbf{0.957~$\pm$~0.026}  &  0.867~$\pm$~0.068  &  0.867~$\pm$~0.068  &  0.872~$\pm$~0.066  &  0.954~$\pm$~0.026 \\
      & T1\textsubscript{fat}  &  \textbf{0.940~$\pm$~0.028}  &  0.866~$\pm$~0.060  &  0.867~$\pm$~0.058  &  0.869~$\pm$~0.057  &  0.937~$\pm$~0.030 \\
      & $\Delta$f                     & \textbf{0.933~$\pm$~0.030}   &  0.824~$\pm$~0.075  &  0.819~$\pm$~0.071  &  0.834~$\pm$~0.069  &  0.919~$\pm$~0.035 \\
      & B1                     & \textbf{0.965~$\pm$~0.018}   &  0.894~$\pm$~0.052  &  0.893~$\pm$~0.052  &  0.895~$\pm$~0.051  &  0.959~$\pm$~0.021 \\
\midrule
R\textsuperscript{2} & FF                     &  \textbf{1.000}  &  0.991  &  0.993  &  0.994  &  0.999 \\
                     & T1\textsubscript{H2O}  &  \textbf{0.919}  &  0.552  &  0.381  &  0.648  &  0.911 \\
                     & T1\textsubscript{fat}  &  \textbf{0.927}  &  0.693  &  0.541  &  0.726  &  0.908 \\
                     & $\Delta$f                     &  \textbf{0.995}  &  0.901  &  0.901  &  0.930  &  0.992 \\
                     & B1                     &  \textbf{0.988}  &  0.865  &  0.785  &  0.897  &  0.979 \\
\bottomrule
\end{tabular}
\end{table*}

\subsection{Blurriness of the Parametric Map Reconstructions}

\begin{figure}[!t]
\centering
\includegraphics[width=0.5\textwidth]{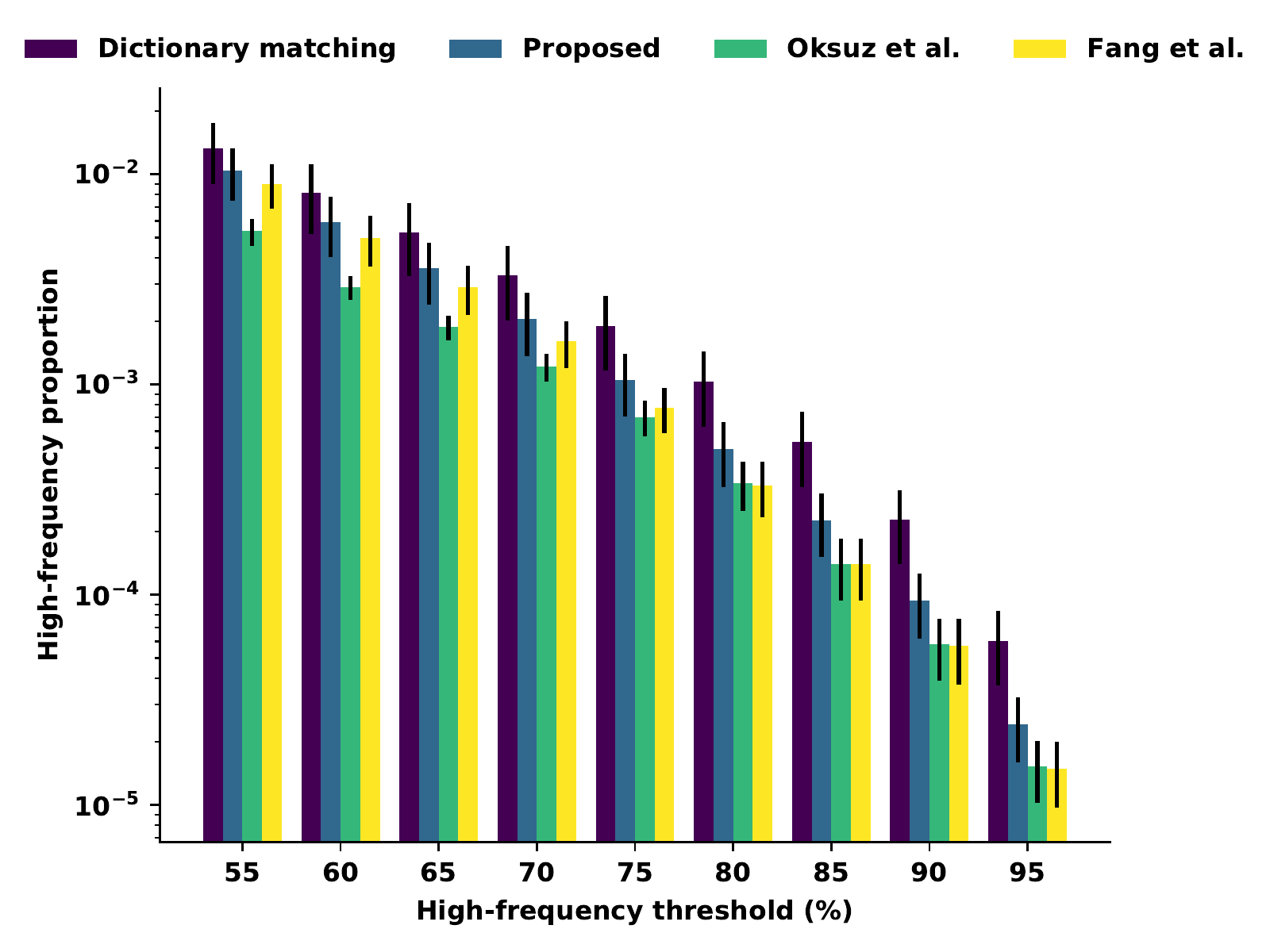}
\caption{Blurriness of the T1\textsubscript{H2O} map reconstruction. The bars indicate mean $\pm$ standard deviation.}
\label{fig:blurriness}
\end{figure}

In a post hoc analysis, we investigated the blurriness (or smoothness) of the reconstructions of the different methods. We analyzed the energy of the high frequencies in the parametric maps as a metric of blurriness, i.e., the ratio between the energy of the high frequencies and the energy of all frequencies (similar to Section~2.1 of the supplementary material of \citet{Fang2019}). We defined the high frequencies in the spectrum of the parametric maps to be the frequencies above a certain threshold, which we varied from 55 to 95~\% because defining one single threshold to separate low and high frequencies was difficult. The energy was defined as the sum of the squared magnitudes. Fig.~\ref{fig:blurriness} compares the blurriness of the T1\textsubscript{H2O} map reconstruction between the methods (see Section~3 of the supplementary material for the FF, T1\textsubscript{fat}, $\Delta$f, and B1 maps). For T1\textsubscript{H2O}, all methods clearly produced smoother reconstructions than the dictionary matching. Further, the visually smoother appearance of the reconstructions of \citet{Fang2019} can be confirmed quantitatively. For the FF and $\Delta$f maps, \citet{Oksuz2019} resulted in less smoothing than the dictionary matching, which also confirms the noisy appearance in Fig.~\ref{fig:qualitative}.

\subsection{Influence of the Spatial Dimension}
\label{sec:spatial}

The influence of the spatial dimension on the reconstruction quality, or in other words, to what extent the correlation of neighboring fingerprints is beneficial for the reconstruction, is to this date not well studied. Therefore, we varied the receptive field of the proposed CNN from 1~$\times$~1, which corresponds to fingerprint-wise reconstruction, up to a receptive field of 21~$\times$~21 using Algorithm~\ref{algo:architecture}. We further varied the number of parameters from 1 to 5 million in steps of 1 million (see Section~4 of the supplementary material for configurations). Fig.~\ref{fig:spatial} shows the R\textsuperscript{2} of the T1\textsubscript{H2O} map reconstruction depending on the receptive field and the number of parameters\footnote{This experiment led to the choice of the final architecture with a receptive field of 15~$\times$~15 and 5~million parameters. Therefore, the results in Fig.~\ref{fig:spatial} are from the validation set because analyzing the test set would violate the independence of architecture design and test set.}. It is visible that fingerprint-wise reconstruction results in significantly inferior reconstructions. Receptive fields around 15~$\times$~15 seem to perform well with little to no added value when incorporating more fingerprints for the reconstruction. There is no significant change in performance with fewer or more number of parameters. The influence on the parametric maps and metrics except for the T1\textsubscript{H2O} map and R\textsuperscript{2} were less accentuated for receptive fields above 5~$\times$~5. Increasing the receptive field beyond 15~$\times$~15 had in some cases negative influence (see Section~4 of the supplementary material). Therefore, we did not experiment with receptive fields beyond 21~$\times$~21. Considering all parametric maps and metrics, we chose a receptive field of 15~$\times$~15 and 5~million parameters. For comparison, we summarize the receptive field and the number of parameters of the methods of comparison in Table~\ref{tab:parameters}.

\begin{figure}[!t]
\centering
\includegraphics[width=0.5\textwidth]{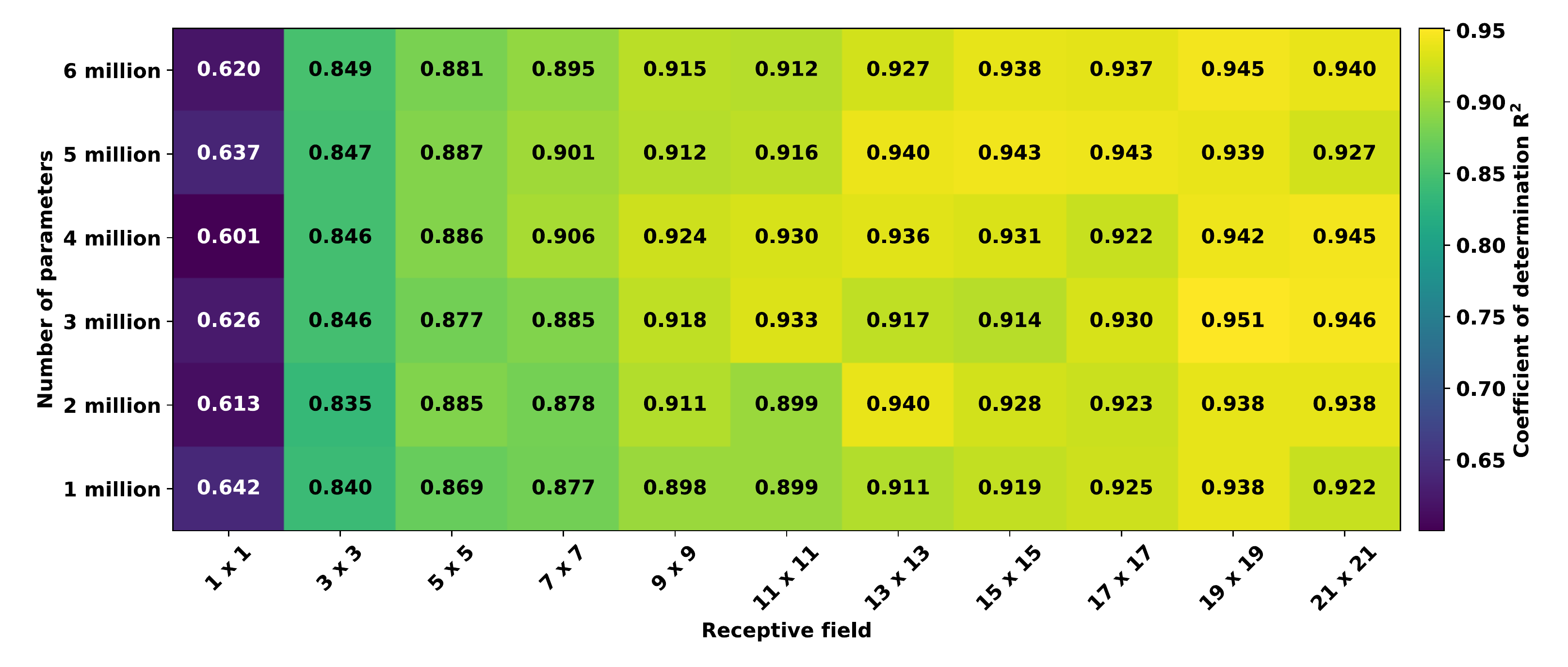}
\caption{Influence of the spatial receptive field and the number of parameters on the T1\textsubscript{H2O} map reconstruction. The numbers denote the R\textsuperscript{2}.}
\label{fig:spatial}
\end{figure}

\begin{table}[!t]
\footnotesize
\renewcommand{\arraystretch}{1.1}
\caption{Architectural summary of the proposed method and the methods of comparison. For \citet{Fang2019} the numbers represent the sum of the numbers of the feature extraction and the spatially-constrained quantification module.}
\label{tab:parameters}
\centering
\begin{tabular}{llll}
\toprule
Method & Number of parameters  & Number of non-linearities & Receptive field \\
\midrule
Proposed & 5.00 million & 21 & 15 $\times$ 15 \\
\citet{Hoppe2017} & 0.05 million & 3 & 1 $\times$ 1 \\
\citet{Cohen2018} & 0.20 million & 2 & 1 $\times$ 1 \\
\citet{Oksuz2019} & 0.12 million & 1 & 1 $\times$ 1 \\
\citet{Fang2019} & 3.84 million & 11 & 54 $\times$ 54 \\
\bottomrule
\end{tabular}
\end{table}

\subsection{Influence of the Temporal Dimension}
\label{sec:temporal}

\begin{figure}[!t]
\centering
\includegraphics[height=0.9\textheight]{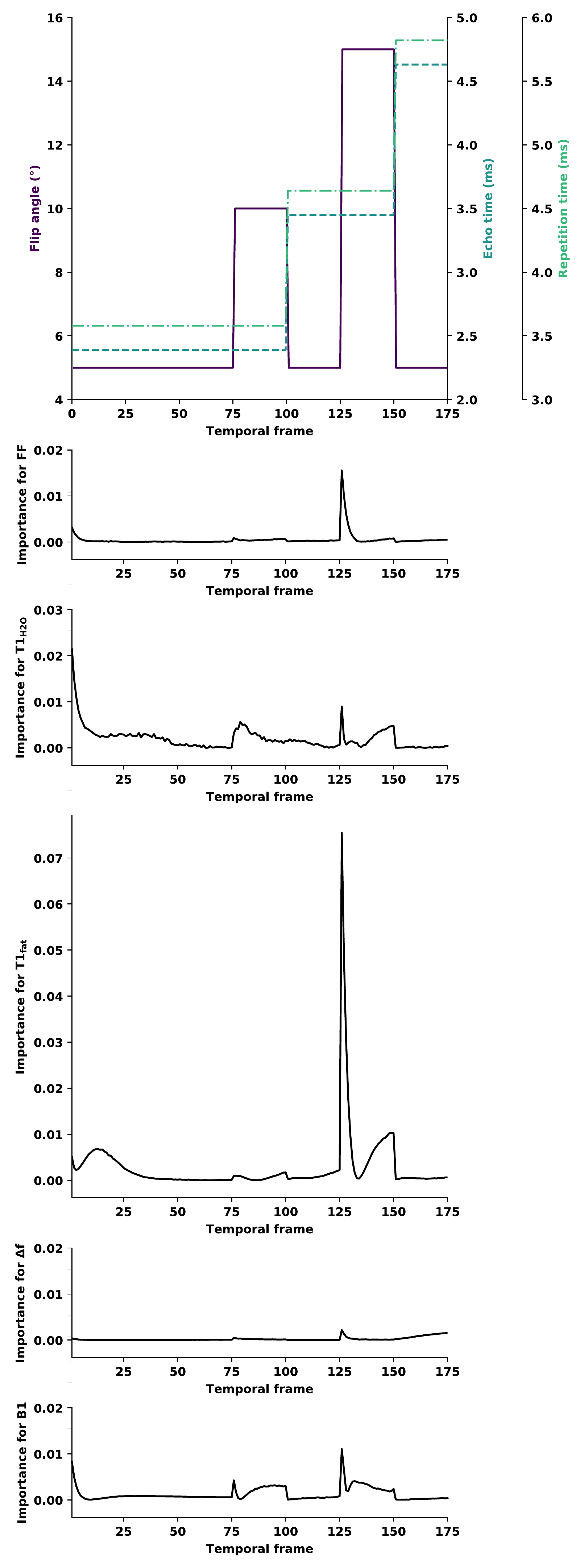}
\caption{Influence of the temporal frames on the parametric map reconstruction. (first row) The MRF T1-FF sequence parameters and the importance of each temporal frame for (second row) FF, (third row) T1\textsubscript{H2O}, (fourth row) T1\textsubscript{fat}, (fifth row) $\Delta$f, and (sixth row) B1 reconstruction.}
\label{fig:temporal}
\end{figure}

The influence of the temporal dimension, or in other words, to what extent the temporal frames contribute to the reconstruction, might be of valuable information for the MRF community. To investigate the influence of the temporal dimension, we reformulated the permutation importance by \citet{Breiman2001,Breiman2001a} for MRF. The permutation importance measures the importance of a variable to a model's prediction accuracy when the variable is permuted \citet{Fisher2019}. Here, the variables are the fingerprint intensities of each temporal frame. We, therefore, randomly permuted the intensities of the $t$-th temporal frame and reconstructed the parametric maps using this permuted MRF data as input. The absolute difference in NRMSE to the non-permuted reconstruction is then considered as the importance of the $t$-th temporal frame of the MRF sequence. The importance of all temporal frames for reconstructing the five parametric maps is shown alongside the MRF sequence in Fig.~\ref{fig:temporal}. The first few temporal frames after the non-selective inversion pulse, which should be sensitive to T1, have the highest importance for the reconstruction of the T1\textsubscript{H2O} map. For T1\textsubscript{fat} the temporal frames after 125 are the most important when water and fat are out of phase. Generally, the temporal frames after the changes in the MRF T1-FF sequence parameters result in high importance for the reconstruction (i.e. at temporal frames 75, 100, 125, and 150). 

\subsection{Robustness to Heterogeneous Morphometric Variations}
\begin{figure*}[!t]
\centering
\includegraphics[width=1.\textwidth]{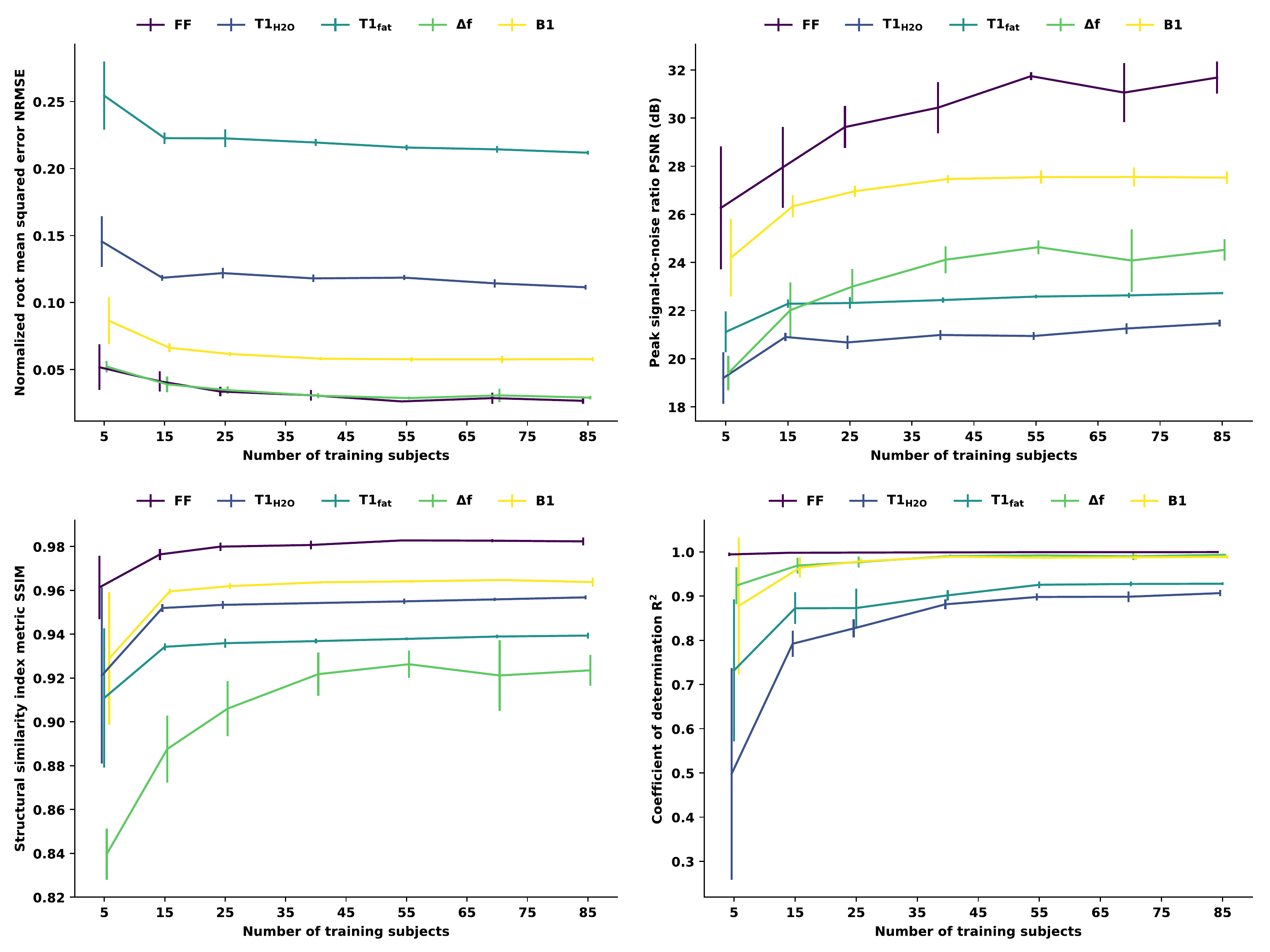}
\caption{Robustness of the proposed method to heterogeneous morphometric variations depending on the number of training subjects. The CNN was trained with 5, 15, 25, 40, 55, 70, and 85 subjects randomly taken from the training split of 94 subjects. The error bars indicate the standard deviation of five random runs.}
\label{fig:robustness}
\end{figure*}

The results show that the proposed method reconstructs highly heterogeneous morphometric variations in NMD patients well. However, it is unclear how many training subjects are actually needed to obtain a model with good robustness. To investigate this, we randomly selected subsets of a varying number of training subjects from the training set, trained the proposed method with these subsets, and reconstructed the testing set to assess the robustness. The whole process was repeated five times. Fig.~\ref{fig:robustness} summarizes the results of this experiment. With 40 training subjects, the proposed method reconstructs almost identically as when using the entire training set with 94 subjects. However, we also observe that the number of required training subjects depends on the metric of interest.

\subsection{Generalization to Unseen Anatomical Regions}
\begin{figure*}[!t]
\centering
\includegraphics[width=1\textwidth]{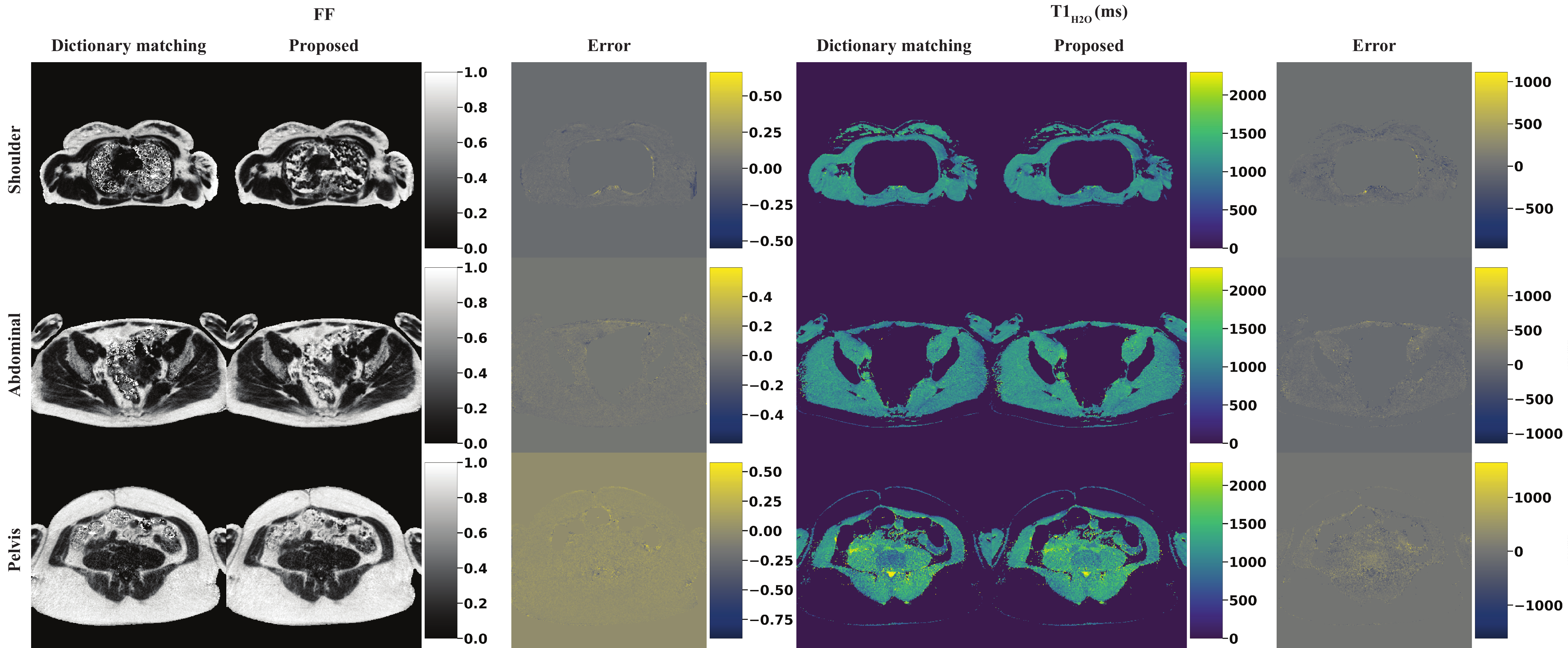}
\caption{Generalization of the proposed method to unseen anatomical regions. The CNN trained on thigh and leg images has been applied to reconstruct shoulder, lower abdominal (proximal to pelvis), and pelvis MRF T1-FF acquisitions. Regions with noisy dictionary matching reconstructions have been masked in the error map due to the insensitivity of MRF T1-FF to tissues other than skeletal muscle and fatty tissue.}
\label{fig:unseen-anatomy}
\end{figure*}

The generalization of deep learning-based MRF reconstruction methods to unseen anatomical regions during training has not been investigated so far, to the best of our knowledge. Therefore, we imaged three NMD patients at three anatomical regions distinctly different to the thigh and leg: the shoulder, the lower abdomen, and the pelvis. MRF acquisition, reconstruction, and evaluation were identical as described in Section~\ref{sec:mrfacquisition} and Section~\ref{sec:evaluation}. FF and T1\textsubscript{H2O} map reconstructions of the proposed method are shown in Fig.~\ref{fig:unseen-anatomy} and the R\textsuperscript{2}s of the ROI analysis for all parametric maps and methods are summarized in Table~\ref{tab:unseen-anatomy} (see Section~5 of the supplementary material for all metrics). Qualitatively, the proposed method reconstructed the parametric maps with good quality. In tissues other than skeletal muscle and fatty tissue, the dictionary matching and the proposed reconstruction resulted in noisy parametric maps, which was expected due to the MRF T1-FF sequence's purpose. Quantitatively, the proposed approach resulted in a slight decrease of performance for FF, $\Delta$f, and B1. For T1\textsubscript{H2O} and T1\textsubscript{fat}, the decrease is more significant (cf. Table~\ref{tab:unseen-anatomy}). This decrease was even more accentuated for the method of \citet{Fang2019}. And, despite working fingerprint-wise, the method of \citet{Oksuz2019} resulted in the worst reconstructions for unseen anatomical regions.

\begin{table}[!t]
\renewcommand{\arraystretch}{1.1}
\caption{Quantitative results for the reconstruction of unseen anatomical regions. The numbers denote the R\textsuperscript{2}.}
\label{tab:unseen-anatomy}
\centering
\begin{tabular}{lccccc}
\toprule
& \multicolumn{5}{c}{Parametric map} \\
\cmidrule{2-6}
Method & FF & T1\textsubscript{H2O} & T1\textsubscript{fat} & $\Delta$f & B1 \\
\midrule
Proposed     & \textbf{0.978} & \textbf{0.862} & \textbf{0.804} & \textbf{0.991} & \textbf{0.989} \\
\citet{Oksuz2019} & 0.746 & 0.409 & 0.420 & 0.862 & 0.760 \\
\citet{Fang2019}  & 0.973 & 0.758 & 0.787 & 0.970 & 0.980 \\
\bottomrule
\end{tabular}
\end{table}

\section{Discussion}

We have investigated the reconstruction of parametric maps from MRF using CNNs. Driven by the hypothesis that the reconstruction performance depends on the incorporation of neighboring fingerprints, i.e., the receptive field of the CNN, and the capacity, i.e., the number of parameters of the CNN, we have designed an algorithm for flexible architecture building based on the specific requirements of the MRF sequence to reconstruct. The configuration for MRF T1-FF was empirically determined to be a receptive field of 15~$\times$~15 with five million parameters. With this configuration, we have shown that the proposed method yields accurate parametric map reconstruction, independent of the morphometric heterogeneity of imaged patients as well as unseen anatomical regions. The method is fast, enabling reconstruction of parametric maps in a clinical setting. Further, as shown qualitatively and quantitatively, better reconstruction results were achieved with the proposed method as compared to other deep learning-based methods.

The proposed method yielded an absolute reconstruction error lower than the dictionary sampling increment for all except the T1\textsubscript{H2O} map (Fig.~\ref{fig:quantitative}). Therefore, we argue that there is no difference in reconstruction accuracy between the proposed method and the dictionary matching for the FF, T1\textsubscript{fat}, $\Delta$f, and B1 maps. Based on the observed differences for T1\textsubscript{H2O} map reconstructions, we concede that the sensitivity of MRF T1-FF to T1\textsubscript{H2O} might not be optimal. The fingerprints encode T1\textsubscript{H2O} to some extent, but for nuances in T1\textsubscript{H2O}, they contain probably more noise than discriminative patterns. This observation can also be confirmed when comparing simulated fingerprints with close T1\textsubscript{H2O} values (see Section~6 of the supplementary material). Further, large receptive fields were mainly beneficial for T1\textsubscript{H2O} with a lower effect on the other parametric maps. The CNN might compensate for the low signal-to-noise ratio by regularizing spatially. We, therefore, also believe that modifications of the CNN architecture will bring limited additional reconstruction performance and that efforts are better invested at optimizing the MRF sequence than the deep learning-based reconstruction such as done recently~\citep{Cohen2017a,Zhao2019,Lee2019a}.

We have analyzed that the receptive field, i.e., considering a spatial neighborhood of fingerprints, influences the reconstruction performance. On the one hand, fingerprint-wise reconstruction (receptive field of 1~$\times$~1) is significantly inferior to spatial reconstruction. We attribute this mainly to the potentially high correlation of neighboring fingerprints coupled with the strong undersampling of MRF. On the other hand, spatial reconstruction has improved the reconstruction only to some extent (Fig.~\ref{fig:spatial}). Regarding the blurriness of the reconstruction, the optimal receptive field is a difficult choice but it lies likely between fingerprint-wise reconstruction and the large receptive field of \citet{Fang2019} (Fig.~\ref{fig:blurriness}). Further, we have observed a larger decrease in performance for \citet{Fang2019} when reconstructing unseen anatomical regions. We attribute this decrease to the method's large receptive field of 54~$\times$~54 where fingerprints without valuable information, possibly influenced by susceptibility artifacts, were included in the reconstruction (cf. noisy tissues in the parametric maps). Therefore, we conclude that pooling operations with subsequent deconvolution operations, as e.g. in the U-Net-like architecture of \citet{Fang2019}, are not needed for MRF reconstruction. Clearly, spatial regularization is superior to fingerprint-wise reconstruction but its extent dependents almost certainly on the MRF sequence due to various factors such as the sensitivity to the MR parameters, \textit{k}-space sampling, in-plane voxel size, among others. But with the proposed algorithm, investigating this aspect becomes straightforward due to the CNN's adaptability.

We have studied the influence of the temporal frames on the parametric map reconstruction. Such reconstruction interpretability might be useful for further developments of MRF reconstruction, as well as the MRF sequence development itself. As expected, the inversion pulse yields high importance to the first few temporal frames for T1 parameters. The general correlation between abrupt sequence parameter changes and high importance hints at highly sensitive temporal frames to MR parameters, and, therefore, rich information for the reconstruction. \citet{Fang2019} proposed to reduce the time of the MRF acquisition by considering only the first fractions of the temporal frames. However, such an approach, although being straightforward, might be suboptimal considering that not all temporal frames might be of equal importance for the MRF reconstruction. For instance, in the case of MRF T1-FF, the temporal frames 50 to 75 as well as the last 25 temporal frames might be useless for the reconstruction, containing maybe redundant or irrelevant information. An acceleration of the MRF acquisition might be achieved by discarding these temporal frames without sacrificing reconstruction performance.

We have demonstrated an excellent robustness of the proposed method to heterogeneous morphometric variations and unseen anatomical regions, a desired key property for image reconstruction~\citep{Knoll2019}. Previous MRF reconstruction studies were performed on small cohorts of healthy volunteers~\citep{Cohen2018,Balsiger2018b,Fang2018,Fang2019,Golbabaee2019,Song2019}, limiting their clinical significance. Here, we have presented, to the best of our knowledge, the first study on reconstructing highly undersampled MRF of patient data only (Table~\ref{tab:materials}). To study the robustness of a method, NMDs are an excellent subject given their large phenotypic variability, the broad range of affected anatomical regions as well as patient age distribution. Our approach seems to be rather insensitive to such variability, which we also attribute to the large and heterogeneous training data. We have found that it is also possible to achieve good robustness with fewer training data (Fig.~\ref{fig:robustness}). However, the robustness is dependent on the parametric map as well as the desired property of the reconstruction (e.g., structural similarity and signal-to-noise). Further, we have found that the number of parameters of the CNN is a rather insensitive characteristic (Fig.~\ref{fig:spatial}). Nor a decrease in performance with fewer neither overfitting with more learnable parameters have been observed, indicating that a performance benefit can primarily be attributed to the spatial regularization (i.e., the receptive field). Reproducing the results of Fig.~\ref{fig:robustness} with varying number of parameters might give additional insights into a possible link between robustness and the number of parameters but is computationally unfeasible due to the immense number of training runs needed.

Our study has several limitations, which we plan to address in future work. First and foremost, the absence of a better reference for comparison than the dictionary matching is a significant issue. Ideally, the CNN reconstruction should be compared to parametric maps obtained by gold standard parametric mapping~\citep{Balsiger2018b}. However, while this is possible for FF (e.g., using 3-point Dixon~\citep{Glover1991}), there exists no MR sequence for T1\textsubscript{H2O} mapping in fatty infiltrated tissue. Second, the optimal MRF data handling is still subject to further research. We have investigated several variants (see Section~1.3 of the supplementary material), but there are certainly open questions such as the complex-valued nature of the MRF data, e.g., reconstruction by complex-valued CNN~\citep{Virtue2017,Trabelsi2018}. Also, the modeling of the temporal domain, e.g., by RNNs as presented by~\citet{Oksuz2019} or by 3-D CNNs, needs further research. Finally, we can not make any statements of the performance on other MRF sequences. Ideally, a completely independent dataset acquired with another MRF sequence and with other diseases should be available to demonstrate applicability among MRF sequences.

In conclusion, we proposed an adaptable CNN for accurate and fast reconstruction of parametric maps from MRF. We demonstrated that incorporating a spatial neighborhood of fingerprints during the reconstruction is beneficial and that we achieved excellent reconstruction accuracy and robustness to heterogeneous patient data. The proposed method could enable MRF beyond clinical research studies.

\section*{Acknowledgments}
This research was supported by the Swiss National Science Foundation (SNSF) under grant number 184273. The authors thank the Nvidia Corporation for their GPU donation and acknowledge the valuable discussions with Pierre-Yves Baudin.

\bibliographystyle{model2-names.bst}\biboptions{authoryear}
\bibliography{library}

\end{document}